\documentclass[twocolumn,amsmath,amssymb,10pt,superscriptaddress,a4paper,letterpaper,fleqn]{revtex4-1}
\usepackage{amssymb}
\usepackage{epsfig}
\usepackage{graphicx}
\usepackage{dcolumn}
\usepackage{array}
\usepackage{bm}
\usepackage{fancyheadings}
\usepackage{longtable}
\usepackage{multirow}
\usepackage{float}
\usepackage{afterpage}
\usepackage{color}
 
\setlongtables
\usepackage[breaklinks=true,linkbordercolor={1 1 1}]{hyperref}
 
\begin{document}\setcounter{page}{1}
 
 
\title{
\qquad \\ \qquad \\ \qquad \\  \qquad \\  \qquad \\ \qquad \\
Towards a coupled-channel optical potential for rare-earth nuclei}
 
\author{G. P. A. Nobre}
\email[Corresponding author: ]{gnobre@bnl.gov}
\affiliation{National Nuclear Data Center, Brookhaven National Laboratory, Upton, NY 11973-5000, USA}
 
\author{A. Palumbo}
\affiliation{National Nuclear Data Center, Brookhaven National Laboratory, Upton, NY 11973-5000, USA}
 
\author{D. Brown}
\affiliation{National Nuclear Data Center, Brookhaven National Laboratory, Upton, NY 11973-5000, USA}
 
\author{M. Herman}
\affiliation{National Nuclear Data Center, Brookhaven National Laboratory, Upton, NY 11973-5000, USA}
 
\author{S. Hoblit} 
\affiliation{National Nuclear Data Center, Brookhaven National Laboratory, Upton, NY 11973-5000, USA}
 
\author{F. S. Dietrich} 
\affiliation{P.O. Box 30423, Walnut Creek, CA, 94598, USA} 
 
\date{\today} 
 
\begin{abstract}
We present an outline of an extensive study of the effects of collective couplings and nuclear deformations on
integrated cross sections as well as on angular distributions in a consistent manner for neutron-induced reactions
on nuclei in the rare-earth region. This specific subset of the nuclide chart was chosen precisely because of a
clear static deformation pattern. We analyze the convergence of  the coupled-channel calculations regarding the
number of states being explicitly coupled. A model for deforming the spherical Koning-Delaroche optical potential as
function of quadrupole and hexadecupole deformations is also proposed, 
inspired by previous works. We demonstrate that the obtained results of calculations for total, elastic, inelastic,
and capture cross sections, as well as elastic and inelastic angular distributions are in  remarkably good agreement
with experimental data for scattering energies around a few MeV. 
\end{abstract}
\maketitle
 
\lhead{Towards a coupled-channel $\dots$}
\chead{NUCLEAR DATA SHEETS}
\rhead{G.~P.~A.~Nobre \textit{et al.}}
\lfoot{}
\rfoot{}
\renewcommand{\footrulewidth}{0.4pt}
 
 
\section{INTRODUCTION}
 
Optical potentials (OP) have been widely used to describe nuclear reaction data by implicitly accounting for the
effects of excitation of internal degrees of freedom and other nonelastic processes. Such optical potentials are
normally obtained through proper parametrization and parameter fitting in order to reproduce specific data sets.
An OP is called global when this fitting process is consistently done for a variety of nuclides.
 
Even though  existing phenomenological OP's  might achieve very good  agreement with experimental data under certain
conditions, as they were specifically designed to do so, they are not reliable at regions without any measurements,
for deformed nuclei, or for the ones away from the stability line. In such circumstances, a more fundamental
approach becomes necessary. 
 
The coupled-channel theory is a natural way of explicitly treating nonelastic channels, in particular those arising
from collective excitations, defined by nuclear deformations. Proper treatment of such excitations is often essential
to the accurate description of reaction experimental data. Previous works have applied different models to specific
nuclei with the purpose of determining angular-integrated cross sections.

There are global spherical OP's that have been fit to nuclei below and above the region of statically deformed
rare-earth nuclei, but these potentials have been viewed as inappropriate for use in coupled-channels calculations,
since they do not account for the loss of flux through the explicitly included inelastic channels.  On the other
hand, a recent paper \cite{Dietrich:2012} shows that scattering from rare earth and actinide nuclei is very near
the adiabatic (frozen nucleus) limit, which suggests that the loss of flux to rotational excitations might be
unimportant.  In this paper we test this idea by performing coupled channel calculations with a global spherical
optical potential by deforming the nuclear radii but making no further adjustments.  We note an alternative
approach (Kuneida \emph{et al.} \cite{Kunieda:2007}), which has attempted to unify scattering from spherical and
deformed nuclei by considering all nuclei as statically deformed, regardless of their actual deformation.
 
\section{Proposed model for rare-earths}
 
To test our proposed model, we deform the widely-used spherical Koning-Delaroche optical potential \cite{KD} and
perform coupled-channel calculations for 34 deformed nuclei, namely, $^{152,154}$Sm, $^{153}$Eu, $^{155,156,157,158,160}$Gd,
$^{159}$Tb, $^{162,163,164}$Dy, $^{165}$Ho,  $^{166,167,168,170}$Er, $^{169}$Tm, $^{171,172,173,174,176}$Yb, $^{175,176}$Lu,
$^{177,178,179,180}$Hf, $^{181}$Ta, and $^{182,183,184,186}$W. For such calculations we used the reaction model code
\textsc{Empire}, which has the direct reaction process calculated by the code \textsc{Ecis}. We then compared the
agreement to experimental data of direct-reaction observables.
 
\begin{figure}[!htb]
 \begin{center}
 \includegraphics[angle=-0.0,scale=0.8,clip=true,trim=94mm 110mm 77.5mm 20mm]{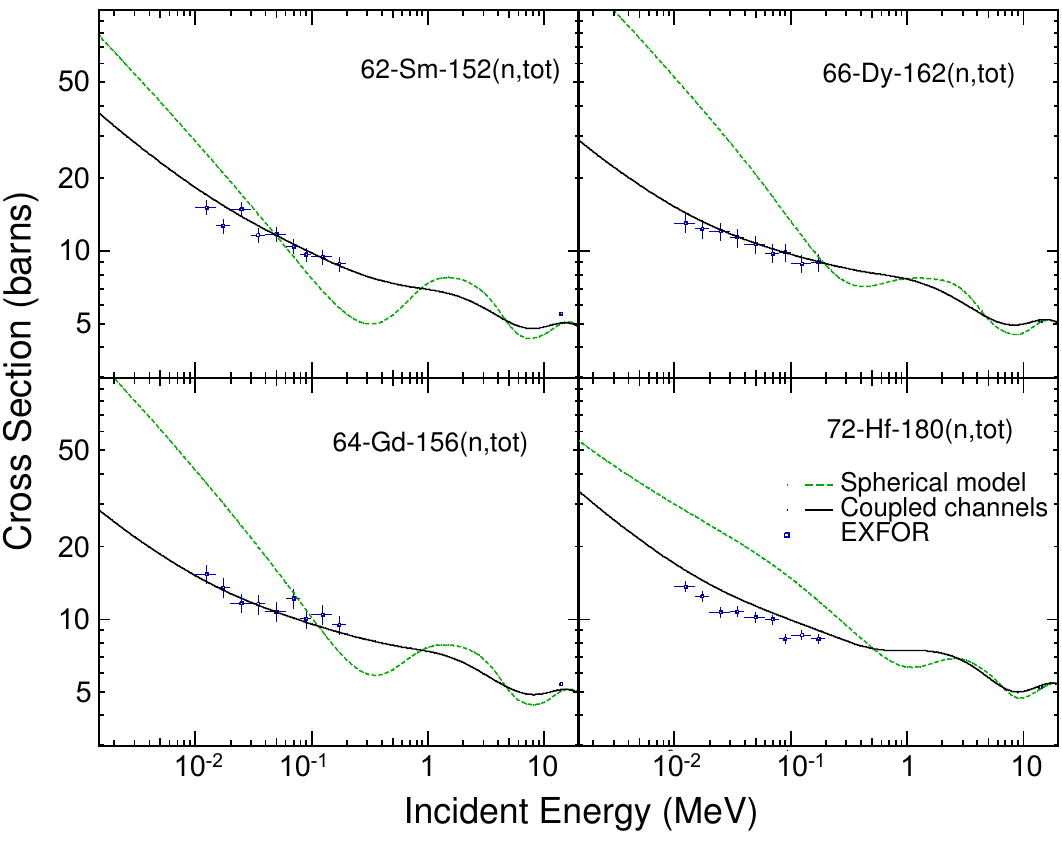}
 \end{center}
 \vspace{-3mm}
 \caption{Total cross sections for the nuclei $^{152}$Sm, $^{156}$Gd, $^{162}$Dy, and $^{180}$Hf
 obtained through deforming the spherical Koning-Delaroche optical potential using coupled channels
 (solid black curves). For comparison purposes the calculation with spherical model is also plotted (green dashed curves).}
 \label{Fig:total}
\end{figure}

Even though the Koning-Delaroche potential is well known to agree very well with existing experimental data for
spherical and stable nuclei, it was not designed to describe the ones in the rare-earth region, due to the high
deformation of such isotopes. Nevertheless, after deforming this potential in the approach of coupled channels, a
good agreement with experimental data for total cross sections is observed, as can be seen in the examples shown in Fig.~\ref{Fig:total}.
 
\subsection{Volume Conservation}
 
\begin{figure}[!htb]
 \begin{center}
 \includegraphics[angle=-0.0,width=1.03\columnwidth,clip=true,trim=5mm 4mm 0mm 5mm]{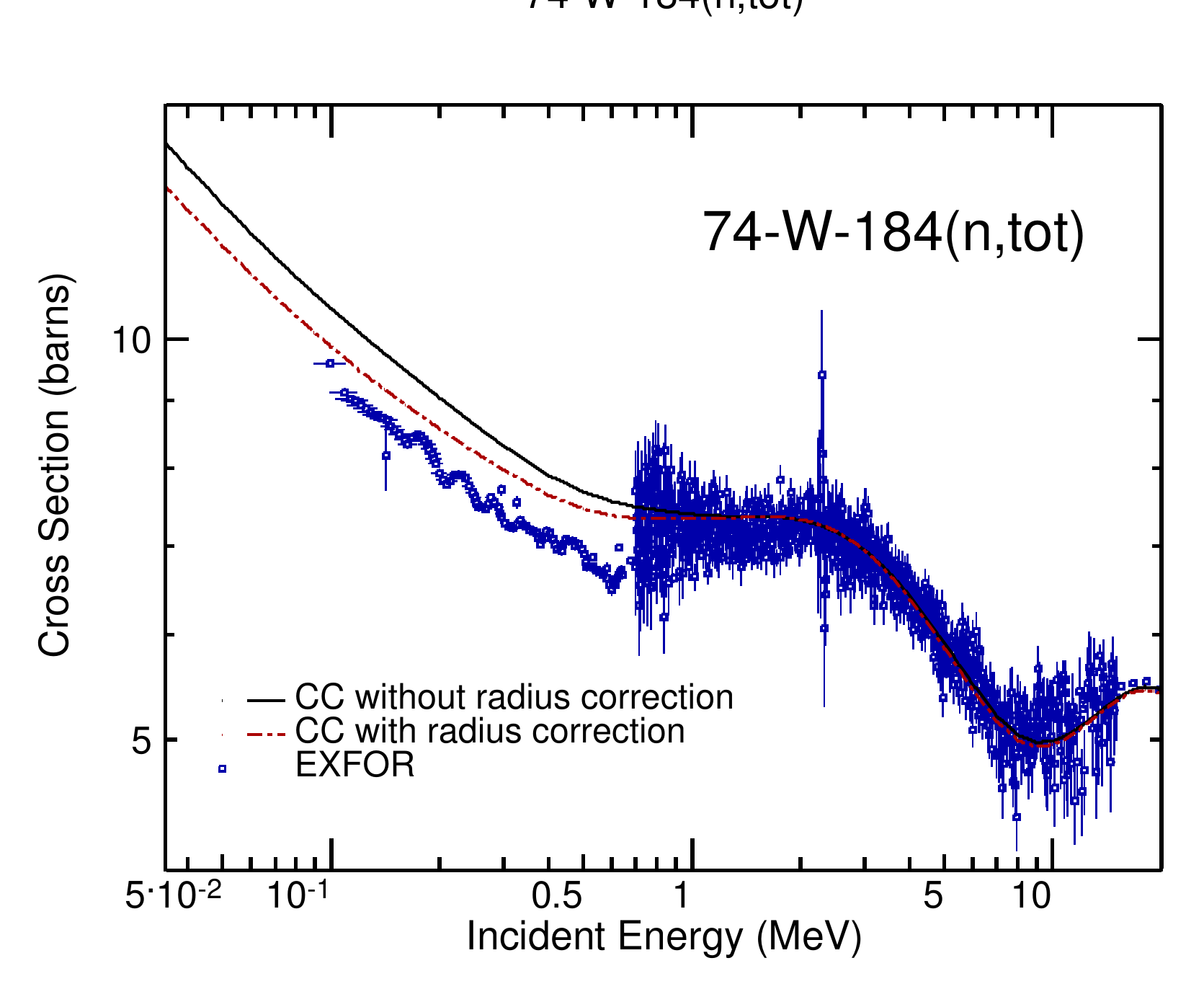} 
 \end{center}
 \vspace{-3mm}
 \caption{Comparison between coupled-channel calculations with (dash-dotted red curve) and without
 (solid black curve) the radius correction to ensure volume conservation (see discussion in text) for $^{184}$W total cross sections.}
 \label{Fig:VolCons-total}
\end{figure}

\begin{figure}[!htb]
 \begin{center}
 \includegraphics[angle=-0.0,width=1.03\columnwidth,clip=true,trim=5mm 4mm 0mm 5mm]{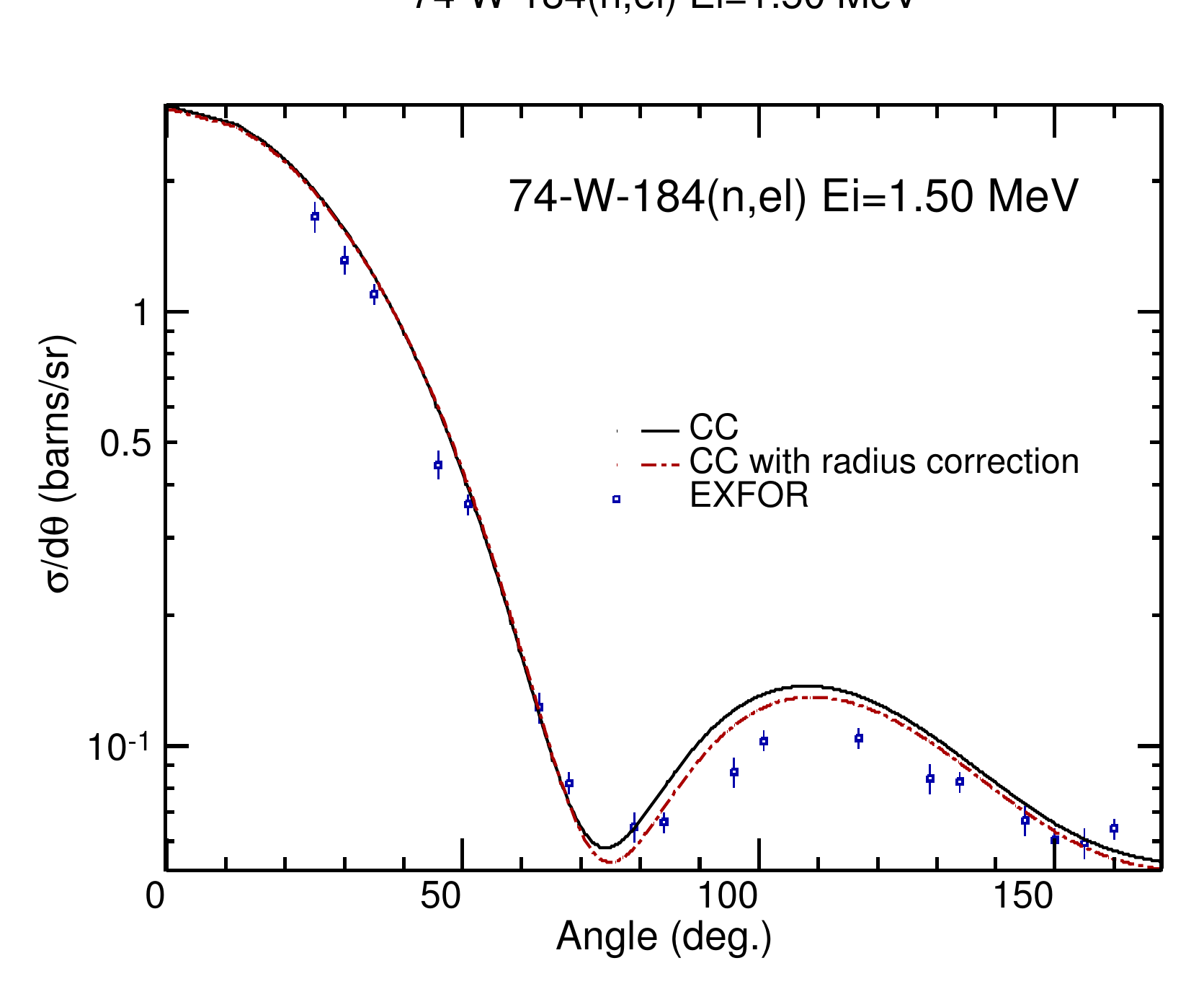} 
 \end{center}
 \vspace{-3mm}
 \caption{Comparison between coupled-channel calculations with (dash-dotted red curve) and without
 (solid black curve) the radius correction to ensure volume conservation (see discussion in text) for $^{184}$W
 elastic angular distribution at the incident energy of 1.50 MeV.}
 \label{Fig:VolCons-angdist}
\end{figure}

When an originally spherical configuration assumes a deformed shape, defined by quadrupole and hexadecupole
deformation parameters $\beta_2$ and $\beta_4$, respectively, the volume and densities are not conserved. In order
to ensure volume conservation we apply a correction to the reduced radius $R_0$, as defined in Ref. \cite{Bang:1980}, which is
\begin{equation}
 R'_{0}=R_0\left(  1-\sum^{}_{\lambda}{\beta_{\lambda}^{2}/4\pi}\right) ,
\label{Eq:radius}
\end{equation}
where terms of the order of $\beta_{\lambda}^{3}$ and higher have been discarded. 
The effect of such correction is not negligible and seems to bring the calculation to a slightly better agreement with
the experimental data, as can be seen in Figs. \ref{Fig:VolCons-total} and \ref{Fig:VolCons-angdist}. This behavior is
observed even though the relatively high deformation of $\beta_2=0.236$ in the case of $^{184}$W, shown in
Figs.~\ref{Fig:VolCons-total} and \ref{Fig:VolCons-angdist}, correspond to a correction in the radius lower than
one half percent. As a matter of fact, of the nuclei studied in this work, the one with highest deformation ($\beta_2=0.353$
for $^{160}$Gd) has a radius correction of less than 1\%.

\subsection{Sensitivity to deformation}
 
\begin{figure}[!htb]
 \begin{center}
 \includegraphics[angle=-0.0,width=1.03\columnwidth,clip=true,trim=5mm 4mm 0mm 5mm]{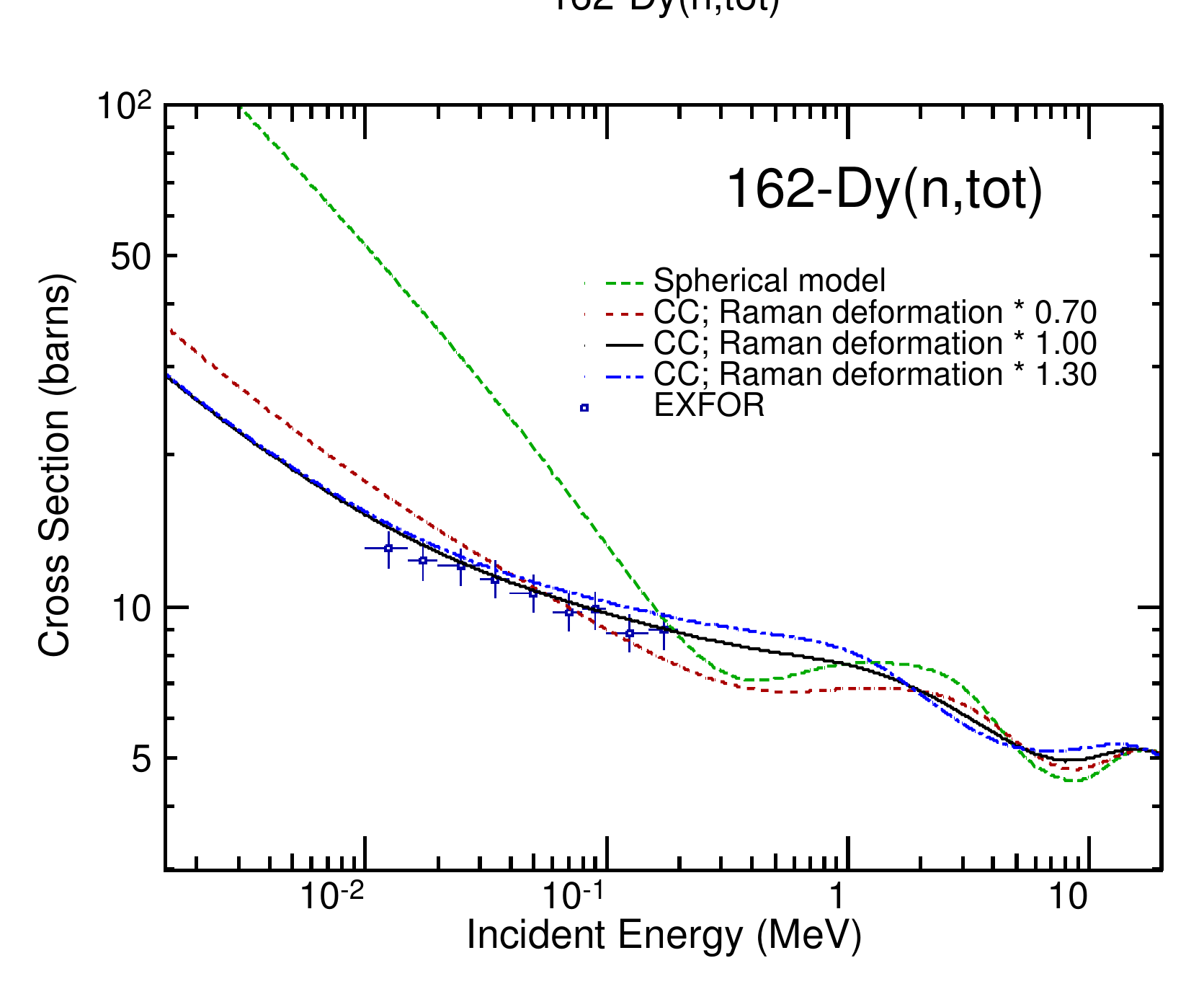} 
 \end{center}
 \vspace{-3mm}
 \caption{Example of the effect of deformation uncertainty in the case of $^{162}$Dy, depicted by the
 results of performing coupled-channel calculations with the value for $\beta_2$ from \cite{Raman:2001} (solid black
 curve) increased (dash-dotted blue curve) and decreased (short-dashed red curve) by 30\%. For comparison purposes
 the calculation with spherical model is also plotted (long-dashed green curve).}
 \label{Fig:DefSens}
\end{figure}

As part of our study we varied the deformation parameters up and down 10\%, 20\%, and 30\%, in relation to the central
adopted value, and we observed the corresponding effect in the direct-reaction observables. As an example, we show in
Fig.~\ref{Fig:DefSens} the total cross-section results for $^{162}$Dy obtained from coupled-channel calculations using
three different values of $\beta_2$: the standard value taken from the compilation of Raman \emph{et al.} \cite{Raman:2001}
multiplied by 0.70, 1.00 and 1.30. For comparison purposes we also plot the result from the spherical model. As can be
seen in Fig. \ref{Fig:DefSens}, for values of  incident energy $E_{\mathrm{i}}$ around $E_{\mathrm{i}} \sim$ 40~keV,
2~MeV, and 5~MeV, the calculated cross sections are insensitive to variations in the deformation.  In the other regions,
however, large deformation uncertainties lead to large cross-section uncertainties. This means that, in order to assess
accurately the value of $\beta_2$, it would be more reasonable to obtain cross-section measurements in energy regions
where this sensitivity is higher. Similar behavior was observed in total elastic cross sections.
 
\section{Angular distributions}
 
\begin{figure}[htb]
 \begin{center}
  \includegraphics[angle=-0.0,width=1.0\columnwidth,clip=true,trim=14mm 34mm 13mm 50mm]{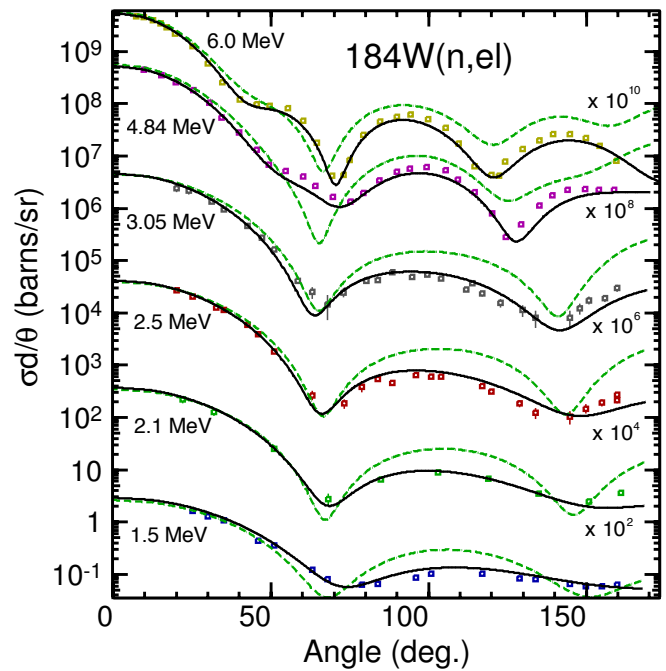} 
 \end{center}
 \vspace{-3mm}
 \caption{Elastic angular distributions for $^{184}$W at different incident energies, which are shown
 on the left-hand side. The solid black curves are predictions of our coupled-channel model and the dashed green curves
 are the spherical model results.}
 \label{Fig:ElasAngDist}
\end{figure}

We also assessed the quality of our model in describing angular distributions. Fig. \ref{Fig:ElasAngDist} shows, as an
example, elastic angular distributions for neutrons scattered by $^{184}$W at different incident energies. As can be
seen in Fig. \ref{Fig:ElasAngDist}, the predictions of our model are in a much better agreement with experimental
data than spherical-model calculations. 
The interesting point to note is that, by deforming the spherical Koning-Delaroche potential, the data description is
not only better than the spherical model, as expected, but also generally very good. This
serves as an indication that the adiabatic limit is a good approach towards an optical potential for the rare-earths.
 
\begin{figure}[htb]
 \begin{center}
 \includegraphics[angle=-0.0,width=0.97\columnwidth,clip=true,trim=48mm 70mm 48mm 82mm]{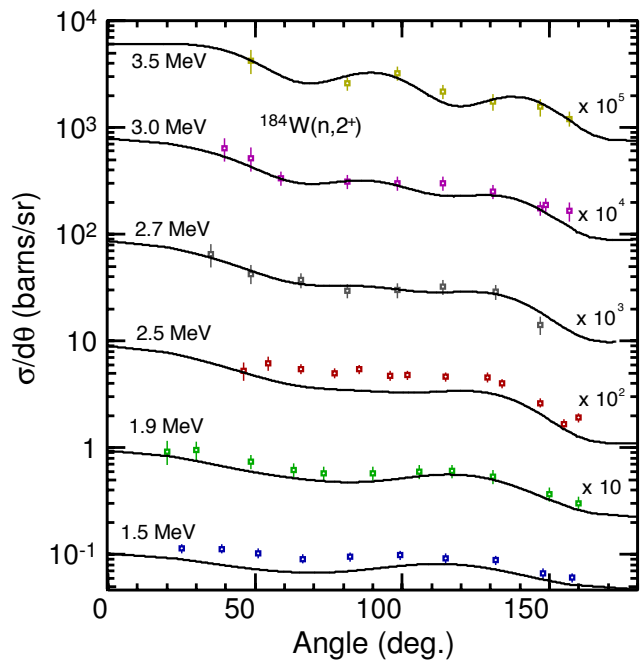} 
 \end{center}
 \vspace{-7mm}
 \caption{Inelastic angular distributions for the 2$^+$ state with excitation energy 111.2~keV 
 of $^{184}$W at different incident energies. The curves are predictions of our coupled-channel model.}
 \label{Fig:InelAngDist}
\end{figure}
 
The same good description of experimental data is observed for inelastic angular distributions. As an example of our
results, we show in Fig. \ref{Fig:InelAngDist} angular distributions for the first 2$^{+}$ state (111.2 keV) of
$^{184}$W for different incident energies. In this case, attempts of comparing to spherical model clearly do not make
sense. Therefore we plot only the predictions of our model. It can be seen in Fig. \ref{Fig:InelAngDist} that,
although the agreement with experimental data is not perfect, the fact that deforming the Koning-Delaroche optical
potential allows to describe inelastic angular distributions reasonably well is very encouraging and supportive of
the model. Subsequent minor adjustments and modifications may be able to account for the existing disagreements.
 
\begin{figure}[htb]
 \begin{center}
 \includegraphics[angle=-0.0,width=1.02\columnwidth,clip=true,trim=5mm 4mm 0mm 5mm]{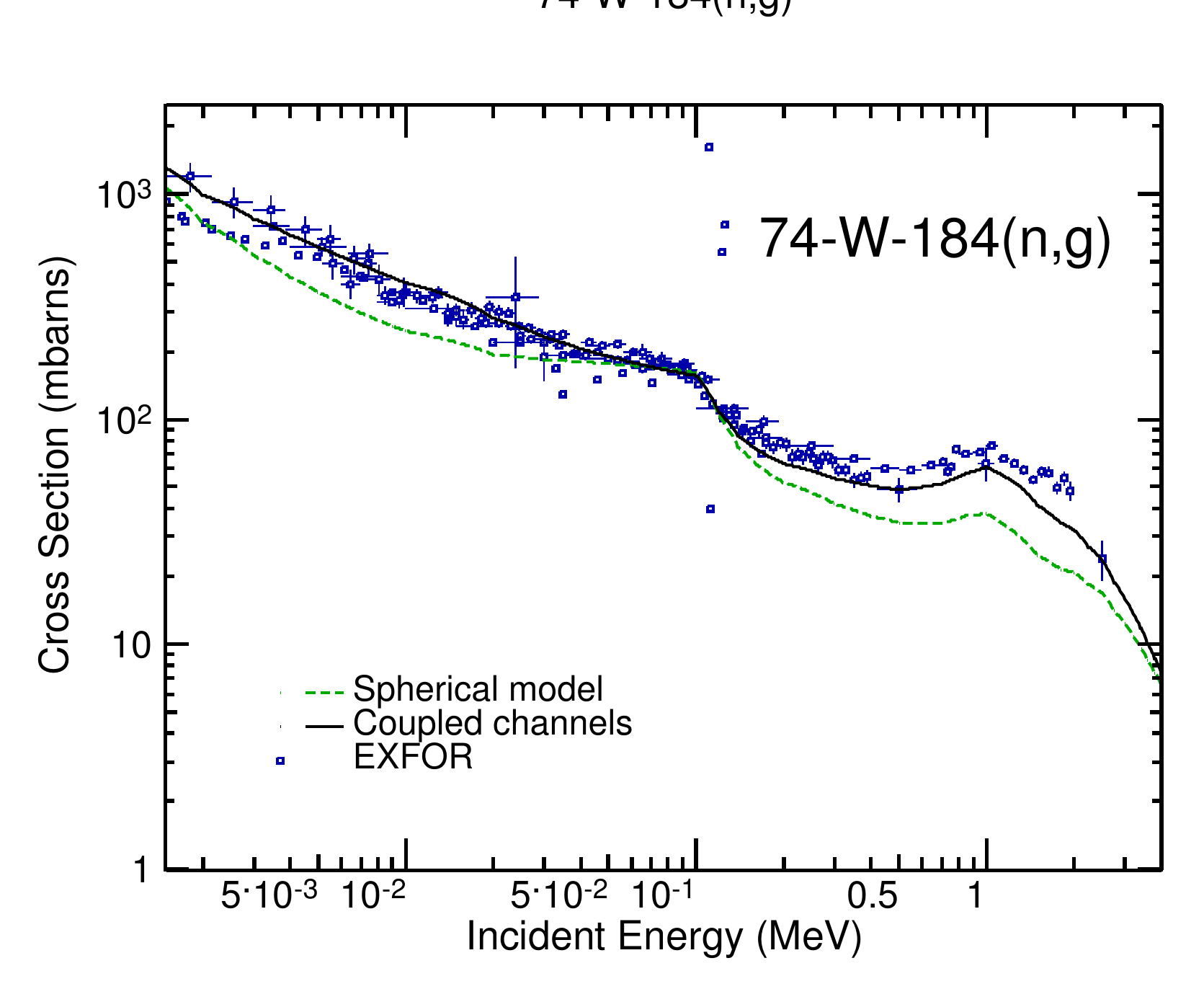} 
 \end{center}
 \vspace{-3mm}
 \caption{Comparison of results for $^{184}$W capture cross sections obtained by our coupled-channel
 (solid black curve) and spherical  (dashed green curve) models.}
 \label{Fig:Capture}
\end{figure}
 
\section{Compound-nucleus observables}
 
The successful results of our approach in describing direct-reaction cross sections served as motivation to test the
effect of our model assumptions on compound-nucleus quantities. We analyzed the results for total elastic, which includes
contributions from shape and compound elastic, total inelastic and capture cross sections. In all cases a significant
improvement in data description was obtained by our coupled-channel model, in comparison to spherical model
calculations. To illustrate such results we plot in Fig. \ref{Fig:Capture} the $^{184}$W(n,$\gamma$) cross sections
for both spherical and coupled-channel models. It is seen that, while the spherical model fails to accurately describe
the shape of observed capture, the cross sections calculated through our coupled-channel approach is in impressive
agreement with experimental data.
 
\section{Conclusion}
 
In this work we have demonstrated that deforming the spherical Koning-Delaroche optical potential 
and using it in coupled channels calculations without further modification provides encouraging
results in the description of neutron-induced reactions on the rare-earths, despite the fact this potential was not
designed to describe such deformed nuclei. We assessed the effect of a correction in the reduced radius to ensure
volume conservation when deforming an originally spherical configuration. This correction was found to produce small
but significant effects in the direction of improving the agreement with experimental data, at least for the cases where tested.
We achieved a good description of experimental data not only for optical-model observables (such as total cross sections,
elastic and inelastic angular distributions), but also for those obtained through compound-nucleus formation (such as
total elastic and inelastic, capture cross sections). These good results are consistent with the insight gained from
Ref. \cite{Dietrich:2012} that the scattering is very close to the adiabatic limit. Although the presented results are
not perfect, this simple method corresponds to a good, consistent and general first step towards an optical potential
capable of fully describing the rare-earth region, filling the current lack of optical model potentials in this important region.

\section*{Acknowledgement}

The work at Brookhaven National Laboratory was sponsored by the Office of Nuclear
Physics, Office of Science of the U.S. Department of
Energy under Contract No. DE-AC02-98CH10886 with
Brookhaven Science Associates, LLC.


\end{document}